\documentclass{iaus}
\usepackage{graphics}

\title[A magnetically collimated jet from W43A] 
{A magnetically collimated jet from the evolved star W43A}

\author[W.~Vlemmings et al.]   
{W.~H.~T.~Vlemmings$^1$, 
 P.~J.~Diamond$^1$ \and H. Imai$^2$}

\affiliation{$^1$Jodrell Bank Observatory, Univ. of Manchester, Macclesfield, Cheshire SK11 9DL, U.K. \break email: wouter@jb.man.ac.uk\\[\affilskip]
$^2$Department of Physics, Faculty of Science, Kagoshima University, Kagoshima 890-0065, Japan}

\pubyear{2006}
\volume{234}  
\pagerange{x--x}
\date{?? and in revised form ??}
\jname{Planetary Nebulae in our Galaxy and Beyond}
\editors{M. J. Barlow \& R. H. M\'endez, eds.}
\begin{document}

\maketitle

\begin{abstract}

We present the first direct measurements of the magnetic field
strength and direction in a collimated jet from an evolved star on its
way to become a planetary nebula. Very Long Baseline Array (VLBA)
observations of the linear and circular polarization of the H$_2$O
masers in the collimated jet of W43A reveal a strong toroidal magnetic
field, indicating that the jet is magnetically collimated. The
magnetic field strength in the jet extrapolated back to the stellar
surface yields a surface field of several Gauss, consistent with the
measurements of maser polarization in a large sample of evolved
stars. The origin of the magnetic field is yet unknown, although the
jet precession might point to the existence of a heavy planet or
stellar companion. This is the first direct observational evidence for
magnetic collimation in the jets, that likely play an important role
in shaping planetary nebulae.
\end{abstract}

\firstsection 
\section{Introduction}

\begin{figure}
 \centerline{
 \scalebox{0.7}{\includegraphics{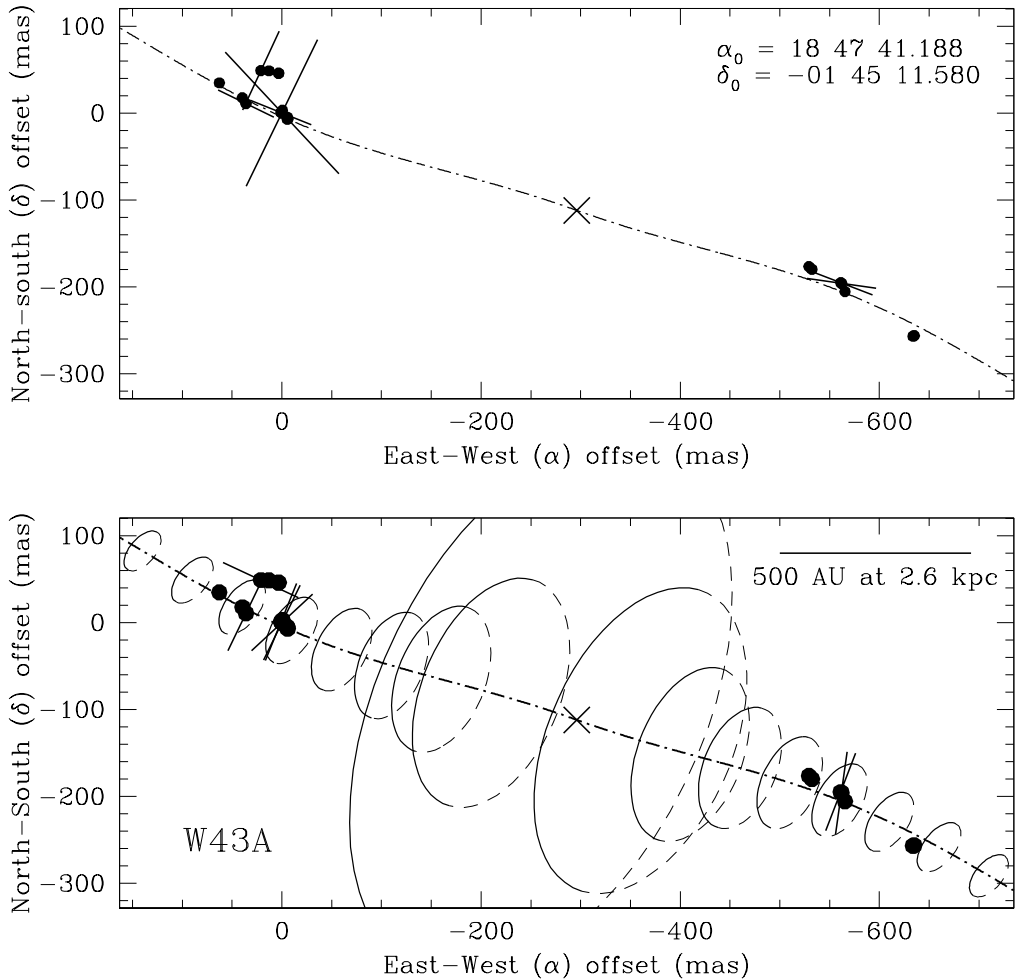}}}
  \caption{The H$_2$O masers in the precessing jet (dashed-dotted
    line) of W43A (indicated by the cross). (top) The maser features
    with the determined linear polarization vectors scaled linearly
    according to the fractional linear polarization. The polarization
    vectors lie predominantly along the jet with a median angle of
    $\chi = 63\pm12^\circ$ east of north. (bottom) The toroidal
    magnetic field of W43A. The vectors indicate the determined
    magnetic field direction, perpendicular to the polarization
    vectors, at the location of the H$_2$O masers. The ellipses
    indicate the toroidal field along the jet, scaled with magnetic
    field strength $\propto r^{-1}$.}
\label{fig:fig1}
\end{figure}

W43A is an evolved star at a distance of 2.6~kpc (\cite[Diamond et
  al. 1985]{Diamond85}). It is surrounded by a thick circumstellar
envelope (CSE) that exhibits OH, H$_2$O and SiO masers (\cite[Imai et
  al. 2005]{Imai05}~and references therein). The H$_2$O masers of W43A
occur in two clusters at $\sim 1000$ AU from the star near the
opposing tips of a collimated jet. The jet, with a velocity of 145
km~s$^{-1}$, has an inclination of 39$^\circ$ with respect to the sky
plane, and a position angle of 65$^\circ$. It precesses by 5$^\circ$
with a period of 55 years. The dynamical age of the jet is inferred to
be only approximately 50 years (\cite[Imai et al. 2002]{Imai02}).
W43A is interpreted as belonging to a class of objects undergoing a
rapid transition from an evolved star into a planetary nebula (PNe).
Currently, owing to their short expected lifetime of less than 1000
years, only 4 sources of this class have been identified (\cite[Imai
  et al. 2002, Morris et al. 2003, Imai et al. 2004, Boboltz \& Marvel
  2005]{Imai02, Imai04, Morris03, Boboltz05}).

Maser polarization observations have revealed the magnetic field
throughout the CSEs of a large number of evolved stars. Close to the
central star, at typically 2 stellar radii, SiO masers indicate
ordered fields of the order of several Gauss (e.g. \cite[Kemball \&
  Diamond 1997, Herpin et al. 2006]{Kemball97, Herpin06}). At the
outer edge of the CSE, the polarization measurements of OH masers
reveal milliGauss magnetic fields and indicate weak alignment with CSE
structure (\cite[Etoka \& Diamond 2004]{Etoka04}). Recently, Zeeman
splitting measurements of H$_2$O masers in the CSEs of a sample of
evolved stars revealed large scale magnetic fields with field
strengths between a hundred milliGauss up to a few Gauss
(\cite[Vlemmings et al. 2002, Vlemmings et al. 2005]{Vlemmings02,
  Vlemmings05b}). While the origin of the magnetic field is still
unclear, theoretical models have shown that a dynamo between the
slowly rotating stellar outer layers and the faster rotating core can
produce the observed magnetic fields (\cite[Blackman et
  al. 2001]{Blackman01}). However, the required additional source of
angular momentum to maintain magnetic field likely requires the
presence of a binary companion or heavy planet (\cite[Blackman et
  al. 2004, Frank et al. 2004, Nordhaus \& Blackman 2006]{Blackman04, Frank04, Nordhaus06}).

Magnetic fields around evolved stars are thought to be one of the main
factors in shaping the CSEs and producing the asymmetries during the
evolution of a spherically symmetric star into the often asymmetric
PNe (Frank 2006, this volume). Similar to well established theories of
collimated outflows in young stellar objects, theoretical models show
that magnetic fields could be the collimating agents of the bi-polar
jets in young proto-planetary nebulae such as W43A
(e.g. \cite[Garc\'ia-Segura et al. 2005]{Garcia05}).  Here we describe
the first direct detection of a magnetically collimated jet from the
evolved star W43A, recently published in \cite{Vlemmings06a}
(hereafter V06a).

\begin{figure}
 \centerline{
 \scalebox{0.4}{\includegraphics{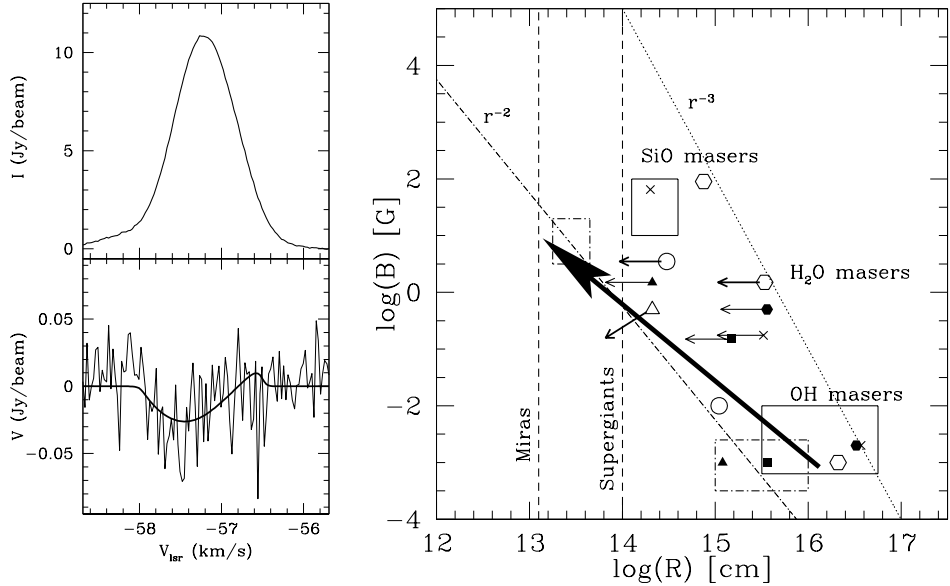}}}
  \caption{(left) The total intensity and circular polarization
    spectrum of the H$_2$O maser feature for which circular
    polarization was detected. (right) The figure reproduced from
    \cite{Vlemmings05b} of measured magnetic fields on the masers in
    the CSEs of evolved stars. The dashed-dotted boxes indicate the
    range of magnetic field strengths measured on the SiO and OH
    masers of Mira stars and the solid boxes those of supergiant
    stars. The thin arrow indicate the H$_2$O maser magnetic fields
    measured in \cite{Vlemmings02, Vlemmings05b}, where the length of
    the arrows indicate the thickness of the H$_2$O maser shell with
    the symbols drawn on the outer edge. The symbols without arrows
    are the measurements on SiO and OH masers from the literature on
    the same sample of stars. The thick solid arrow indicates the
    magnetic field in the jet of W43A.}
\label{fig:spec}
\end{figure}

\section{Observations}

The observations of W43A were performed with the NRAO\footnote{The
  National Radio Astronomy Observatory (NRAO) is a facility of the
  National Science Foundation operated under cooperative agreement by
  Associated Universities, Inc.} VLBA on December 8 2004 at the
frequency of the $6_{16} - 5_{23}$ rotational transition of H$_2$O,
22.235080 GHz. The correlation and data-reduction was performed as
described in \cite{Vlemmings06}, where similar methods were used to
study H$_2$O maser magnetic fields in a high-mass star-forming
region. Polarization calibration was performed with respect to the
calibrator J1743-0350 and we estimate the systematic error of the
polarization angles to be at most $\sim$8$^\circ$.

\section{Results}

Linear polarization was determined on several 22~GHz H$_2$O maser
features in both the red-shifted and blue-shifted tip of the jet of
W43A.  The linear polarization vectors are either parallel or
perpendicular to the magnetic field (\cite[Goldreich et
  al. 1973]{Goldreich73}). As discussed in V06a and elaborated upon in
Vlemmings et al. (in preparation), we find that the polarization
vectors of the masers of W43A are mostly perpendicular to the magnetic
field direction. The observed maser features and their linear
polarization vectors are shown at the top of Fig.\ref{fig:fig1} and
the derived magnetic field direction is shown in the bottom panel of
the figure. In addition to the linear polarization we detected
circular polarization of $P_V = 0.33 \pm 0.09\%$ (Fig.~\ref{fig:spec},
left), which as shown in V06a indicates a deprojected magnetic field
strength of $B=200 \pm 78$~mG.

\section{Discussion}

As the H$_2$O masers are excited in swept up material in the jet, the
magnetic field in the maser region is enhanced. Partial coupling of
the magnetic field to the gas indicates the toroidal magnetic field
strength around the collimated jet, at $\sim 1000$~AU from W43A is
$B_\phi\sim 2$~mG (V06a). The linear polarization vectors,
perpendicular to the jet axis, indicate that we are tracing a
confining toroidal magnetic field. A toroidal field depends on
distance from the star $r$ as $B_\phi \propto r^{-1}$ and we thus find
that at the surface of the star, the magnetic field strength is
between $\sim 2$-$20$~G. As shown in Fig.~\ref{fig:spec}(right) this
is fully consistent with earlier measurements of the magnetic field in
the envelopes of evolved stars. SiO maser observations (\cite[Imai et
  al. 2005]{Imai05}) seem to indicate that the jet formation occurs
within the SiO maser region, close to the star, and thus that the collimating agent is related to the star itself. Considering the shape
and strength of the magnetic field in the jet of W43A, we conclude
that the magnetic field is the primary collimating agent, making this
the first direct detection of magnetic collimation in any
astrophysical jet.

\begin{acknowledgements}
The work by WV has been supported by a Marie Curie Intra-European fellowship
within the 6th European Community Framework Program under contract
number MEIF-CT-2005-010393.
\end{acknowledgements}

\end{document}